\documentclass[10pt,conference]{IEEEtran}
\IEEEoverridecommandlockouts

\usepackage{ifthen}
\usepackage{xcolor}
\usepackage{mathabx}

\usepackage{graphicx}

\usepackage{makecell}

\newcolumntype{D}{ >{\arraybackslash} m{7cm} }
\newcolumntype{C}{ >{\centering\arraybackslash} m{1cm} }
\usepackage{colortbl}
\definecolor{cellorange}{rgb}{ 1,  .949,  .8}
\definecolor{cellgreen}{rgb}{ .776,  .878,  .706}
\definecolor{cellblue}{rgb}{ .608,  .761,  .902}

\usepackage{url}
\usepackage{balance}
\begin{document}

\newboolean{hidecomments}
\setboolean{hidecomments}{false}

\ifthenelse{\boolean{hidecomments}}
{\newcommand{\nb}[2]{}}
{\newcommand{\nb}[2]{
    \fbox{\bfseries\sffamily\scriptsize#1}
    {\sf\small$\blacktriangleright$ {#2} $\blacktriangleleft$}}} 

\newcommand\todo[1]{\nb{TODO}{\textcolor{purple}{#1}}}
\newcommand\done[1]{\nb{DONE}{\textcolor{OliveGreen}{#1}}}

\title{Smart Contract Security: a Practitioners' Perspective
}

\author{\IEEEauthorblockN{Zhiyuan Wan\IEEEauthorrefmark{1}\IEEEauthorrefmark{2}\thanks{\IEEEauthorrefmark{2}Also with PengCheng Laboratory.}, Xin Xia\IEEEauthorrefmark{3}\IEEEauthorrefmark{6}\thanks{\IEEEauthorrefmark{6}Corresponding author.}, David Lo\IEEEauthorrefmark{4}, Jiachi Chen\IEEEauthorrefmark{3}, Xiapu Luo\IEEEauthorrefmark{5}, Xiaohu Yang\IEEEauthorrefmark{1}}
\IEEEauthorblockA{\IEEEauthorrefmark{1}College of Computer Science and Technology, Zhejiang University\\
\IEEEauthorrefmark{3}Faculty of Information Technology, Monash University\\
\IEEEauthorrefmark{4}School of Information Systems, Singapore Management University\\
\IEEEauthorrefmark{5}Department of Computing, Hong Kong Polytechnic University \\
\{wanzhiyuan,yangxh\}@zju.edu.cn, \{xin.xia,jiachi.chen\}@monash.edu, davidlo@smu.edu.sg, csxluo@comp.polyu.edu.hk}
}

\maketitle

\begin{abstract}
Smart contracts have been plagued by security incidents, which resulted in substantial financial losses. Given numerous research efforts in addressing the security issues of smart contracts, we wondered how software practitioners build security into smart contracts in practice.
We performed a mixture of qualitative and quantitative studies with 13 interviewees and 156 survey respondents from 35 countries across six continents to understand practitioners' perceptions and practices on smart contract security. Our study uncovers practitioners' motivations and deterrents of smart contract security, as well as how security efforts and strategies fit into the development lifecycle. We also find that blockchain platforms have a  statistically significant impact on practitioners' security perceptions and practices of smart contract development. Based on our findings, we highlight future research directions and provide recommendations for practitioners.
\end{abstract}
\section{Introduction}\label{sec:introduction}
Blockchain is a distributed ledger that provides an open, decentralized, and fault-tolerant transaction mechanism. Blockchain technology has attracted considerable attention from both industry and academia since it is originally introduced for Bitcoin~\cite{nakamoto2008bitcoin} to support the exchange of cryptocurrency. Blockchain technology evolves to facilitate general-purpose computations with a wide range of decentralized applications. The \emph{Smart contract} technology is one appealing decentralized application that enables the computations on top of a blockchain.

A smart contract is a piece of executable code that runs on a blockchain to enforce the terms of an agreement between untrusted parties. 
Blockchain technology assures that a smart contract is immutable and contract initiated transactions are autonomously and truthfully executed.    
There are multiple blockchain platforms that support smart contracts~\cite{wan2019programmers}, e.g., Ethereum, Hyperledger Fabric, and Corda, with Ethereum being the most prominent  platform~\cite{alharby2017blockchain}.

During the last decade, smart contracts have been plagued by security incidents, which led to losses reaching millions of dollars~\cite{tsankov2018securify}. In June 2016, an attacker exploited vulnerabilities in the DAO smart contract to empty out around 4 million Ethers (worth around 50 million dollars). In July 2017, over 150 thousand Ethers (worth over 34 million dollars) had been stolen due to an exploit in widely-used Parity Wallet~\cite{parity2017jul}. In November 2017, over 500 thousand Ethers (worth over 150 million dollars) were frozen due to a vulnerability in the very same wallet~\cite{parity2017nov}.

To address the security issues of smart contracts, researchers have proposed a broad range of defense solutions, including language-based security~(e.g., \cite{vyper,bamboo,coblenz2017obsidian}), static analysis~(e.g., \cite{luu2016making,tsankov2018securify,tikhomirov2018smartcheck}), and runtime verification~(e.g., \cite{rodler2019sereum}). Vyper~\cite{vyper} removes some of the language functionalities in Solidity to eliminate vulnerabilities and adds new features to support security and readability. In terms of static analysis, Oyente~\cite{luu2016making} leverages symbolic execution to traverse feasible execution paths on control flow graphs and detect vulnerabilities in smart contracts; Securify~\cite{tsankov2018securify} defines compliance and violation patterns based on known vulnerabilities; SmartCheck~\cite{tikhomirov2018smartcheck} translates smart contract code into an XML-based parse-tree and check the parse-tree against specific XPath patterns. With respect to runtime verification, Sereum~\cite{rodler2019sereum} uses taint analysis to monitor runtime data flows during the execution of smart contracts for preventing re-entrancy attacks.

Despite numerous efforts in assuring the security of smart contracts, little is known about how software practitioners build security into smart contracts in practice.
Thus, we followed a mixed methods approach to investigate the practitioners' perceptions and practices with respect to smart contract security. We started with semi-structured interviews with 13 software practitioners with experience in smart contract development, who have an average of 6.5 years of software professional experience. Through the interviews, we qualitatively investigated the security awareness and practices that our interviewees experienced in smart contract development. We derived 6 competing priorities in smart contract development, a list of 11 security motivators\footnote{\emph{Security motivators} are the factors that motivate practitioners to address security; on the contrary, \emph{security deterrents} are the factors that deter practitioners from devoting efforts to security~\cite{assal2019think}.} and 9 security deterrents for smart contract practitioners, 5 sources where practitioners acquire security knowledge, 11 security strategies and 11 factors that affect the adoption of security tools. We further performed an exploratory survey with 156 smart contract practitioners from 35 countries across six continents to quantitatively validate the security perceptions and practices that are uncovered in our interviews. The survey respondents work on multiple blockchain platforms, i.e., public blockchains (80), consortium blockchains (49), and private blockchains (20), and hold various job roles, i.e., development (130), testing (3), and project management (16). We investigated the following research questions:

\vspace{0.1cm}
\noindent\textbf{RQ1. What are practitioners' perceptions regarding smart contract security?}

85\% and 69\% of the survey respondents perceive the importance of security and privacy in smart contracts, respectively. The security motivators include practitioners' awareness of importance, workplace environment, and perceived negative consequences of security issues. Meanwhile, the security deterrents include competing priorities in smart contract development and no formal process to address smart contract security.

\vspace{0.1cm}
\noindent\textbf{RQ2. How does security fit into the development lifecycle of smart contracts?}

This research question investigates security efforts, security strategies, and the adoption of security tools in smart contract development. On average, security efforts account for 29\% of the overall efforts during the development process of smart contracts. To ensure smart contract security, practitioners distribute efforts across different stages in the development lifecycle. They tend to spend significantly more effort towards security at the construction and testing stages than at other stages. In terms of security strategies and tool adoption, 72\% of the respondents frequently leverage more than one security strategy. 58\% of the respondents frequently used the code reuse strategy. 54\% of the respondents frequently adopt security tools, especially the security plugins in integrated development environments.

\vspace{0.1cm}
\noindent\textbf{RQ3. Do blockchain platforms influence practitioners' perceptions and practices on smart contract security?}

Blockchain platforms significantly impact security perceptions and practices of practitioners in smart contract development, including security motivators (e.g., the immutability of smart contracts), security deterrents (e.g., the pressure of feature delivery), the amount of security efforts across stages in the development lifecycle (e.g., security efforts at the construction stage), and strategies to address smart contract security (e.g., code review).

Based on the findings, we discuss the disconnect between the high awareness of smart contract security of practitioners and the frequent occurrence of security problems in smart contracts. We also provide practical lessons about code reuse, tool implications, and proactive defense to ensure smart contract security. In addition, we highlight several research avenues across blockchain platforms.

This paper makes the following contributions:
\begin{itemize}
    \item We perform a mixture of qualitative and quantitative studies to investigate the security perceptions and practices in smart contract development;
    \item We provide practical implications for practitioners and outlined future avenues of research.
    \item We provide the interview guide, questionnaire, and survey responses publicly accessible for future investigation by others\footnote{http://doi.org/10.5281/zenodo.4005112}.
\end{itemize}

The remainder of the paper is structured as follows. Section~\ref{sec:related} briefly reviews  related work. In Section~\ref{sec:methodology}, we describe the methodology of our study in detail. In Section~\ref{sec:result}, we present the results of our study. We discuss the implications of our results in Section~\ref{sec:discussion} and threats to the validity of our findings in Section~\ref{sec:threat}. Section~\ref{sec:conclusion} draws conclusions and outlines avenues for future work. 
\begin{figure*}
    \center
    \includegraphics[scale=0.55]{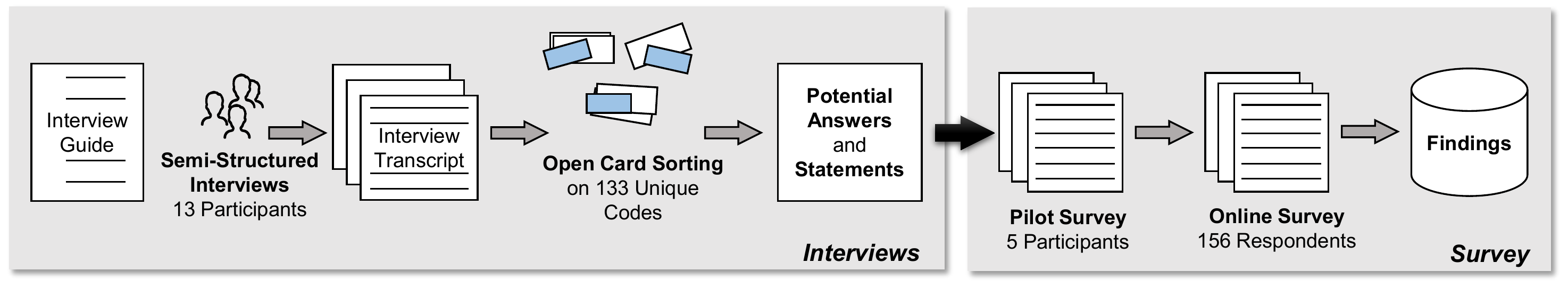}\vspace{-0.2cm}
    \caption{Research methodology.}
    \label{fig:process} \vspace{-0.5cm}
    \end{figure*}
\section{Related Work}\label{sec:related}
\subsection{Security Practices in Software Development}
Practitioners work within organizations, teams, communities, and cultures. Previous studies investigated the social factors that could impact various aspects of security practices, e.g., security tool adoption~\cite{30xiao2014social,32witschey2015quantifying,wan2018perceptions}. Organization and team policies are a driving factor to tool adoption~\cite{58christakis2016developers}, though many organizations do not encourage the adoption of security tools. Large organizations make more use of security tools than small ones~\cite{30xiao2014social}. Existing tools fail to meet the expectations of practitioners by generating low-quality warning messages~\cite{58christakis2016developers}, interrupting work flow~\cite{52johnson2013don,58christakis2016developers,60thomas2016questions}, producing excessive false positives~\cite{52johnson2013don,58christakis2016developers}, and integrating poorly with Integrated Development Environments (IDEs)~\cite{52johnson2013don}.
In this work, we investigated the factors that impact security practices in smart contract development.

Security is expected to be included in developing high-quality software systems, but is rarely listed as an explicit requirement~\cite{35poller2017can}. Practitioners prioritize functional requirements over security and focus on tasks for which outcomes are easy to measure~\cite{30xiao2014social,35poller2017can,37assal2018security}. Pressures from budget and deadlines can also lead to lowering the priority of security practices~\cite{29xie2011programmers}. Some organizations attempt to use penetration testing to motivate practitioners, but the motivation is hard to sustain without continuous support~\cite{33turpe2016penetration}. 
In this work, we investigated how practitioners prioritize security in smart contract development.

Building security in from the start requires a large amount of knowledge. Weir et al. found that enthusiasm about security and motivation to learn are more likely to drive the acquisition of security knowledge for developers than task driven~\cite{34weir2016improve}. Alternatively, security experts could act as a roving source of security knowledge, but face challenges to convince others of the importance of security and examine all generated code with limited resources~\cite{38thomas2018security}. 
Practitioners leverage various information sources to gain knowledge on code security, e.g., documentation~\cite{69nadi2016jumping,71acar2017comparing}, and Stack Overflow~\cite{11acar2016you}. Acar et al.~\cite{11acar2016you} conducted an empirical study to investigate how the use of information sources impacts code security. They found that developers who use Stack Overflow are more likely to produce functional code, but less likely to write secure code. 
This paper investigates the involvement of security experts in smart contract development and information sources of smart contract security.

In the course of software design and construction, misuse of application programming interfaces (APIs) can introduce security vulnerabilities~\cite{ICSE-14-egele2013empirical,ICSE-15-fahl2012eve,ICSE-16-georgiev2012most}. Developers incorrectly use an API because they do not conduct an additional check but trust the API to do the right thing~\cite{73oliveira2018api}. Acar et al. compared the usability of five Python cryptographic APIs and suggested that documentation with examples is more helpful than a simple API~\cite{71acar2017comparing}. Nadi et al.~\cite{69nadi2016jumping} performed an empirical study on how developers use Java cryptography APIs. They found that developers struggle with Java cryptography APIs and prefer task-based solutions. In addition, developers have difficulty in using security-related APIs, for instance, APIs in Transport Layer Security (TLS) and Security Sockets Layer (SSL)~\cite{65fahl2013rethinking,68oltrogge2015pin}. This paper investigates whether the use of smart contract related APIs may introduce security risks.

\subsection{Smart Contract Security}

Security vulnerabilities spread across smart contracts of various blockchain platforms, e.g., Ethereum~\cite{jiang2018contractfuzzer,luu2016making,rodler2019sereum}, Hyperledger Frabric~\cite{yamashita2019potential} and EOS~\cite{quan2019evulhunter}. Security vulnerabilities result from multiple causes, e.g., reentrancy~\cite{rodler2019sereum}, delegatecall injection~\cite{qureshi2017hacker}, and integer overflow and underflow~\cite{integer_cve}. Different programming languages of smart contracts and blockchain architectures lead to different vulnerabilities~\cite{chen2020survey}. In this work, we investigate practitioners' awareness of security vulnerabilities in their smart contracts.  

Previous studies proposed a wide range of approaches and tools for securing smart contracts, including recommending best practices in programming smart contracts, implementing specific programming languages, static analysis, and runtime monitoring. For instance, ConsenSys~\cite{diligence2018ethereum} provides extensive best practices for Ethereum smart contract security, including code patterns to learn and pitfalls to avoid. \emph{Vyper}~\cite{vyper} and \emph{Bamboo}~\cite{bamboo} provide language-based support to eliminate smart contract vulnerabilities. Static analysis tools leverage symbolic execution~\cite{ consensys2019Mythril,chang2019scompile,krupp2018teether,luu2016making,nikolic2018finding,permenev2020verx}, abstract interpretation~\cite{grech2018madmax,kalra2018zeus,suiche2017porosity,tsankov2018securify,zhou2018erays}, formal verification~\cite{grishchenko2018foundations,amani2018towards,grishchenko2018semantic,hirai2017defining,hildenbrandt2018kevm,park2018formal}, fuzzing~\cite{liu2018reguard,jiang2018contractfuzzer,wang2019superion} and model checking~\cite{tikhomirov2018smartcheck} to identify smart contract vulnerabilities. DappGuard~\cite{cook2017dappguard} and Sereum~\cite{rodler2019sereum} monitor the runtime execution of a smart contract to prevent potential exploitations of vulnerabilities. In this work, we investigate the adoption of security strategies and tools of smart contracts in practice and explore the expectations of practitioners. \section{Methodology}\label{sec:methodology}
Our research methodology followed a mixed methods approach~\cite{creswell2017research}, as depicted in Fig.~\ref{fig:process}. The approach follows a sequential explanatory strategy, involving two phases -- a first qualitative phase of interviews, followed by a second quantitative phase of an exploratory survey\footnote{The interviews and survey were approved by the relevant institutional review board (IRB). Participants were instructed that we wanted their opinions; privacy and sensitive resources were not explicitly mentioned}. The survey builds on the results of the interviews. Specifically, we collected data from two sources: (1) We interviewed 13 software practitioners with experience in smart contract development and derived a list of statements and potential answers for survey questions from the results of interviews; (2) We surveyed 156 respondents with experience in smart contract development. To preserve the anonymity of participants, we anonymized all items that constitute of Personally Identifiable Information (PII) before analyzing the data, and further considered aliases as PII throughout our study (e.g., refer to the interviewees as P1 - P13).

\subsection{Interviews}
\label{sec:interview}
The left part of Fig.~\ref{fig:process} describes the process of interviews.
\subsubsection{Protocol}
The first author conducted a series of face-to-face interviews with 13 software practitioners with experience in smart contract development. 
Each interview took 30-45 minutes.
The interviews were semi-structured and made use of an \emph{interview guide}\footnote{Interview guide online: \url{http://doi.org/10.5281/zenodo.4005112}}. To develop the interview guide, we obtained an initial set of open-ended questions through brainstorming within the authors of this paper, focusing on practitioners’ perceptions and practices concerning smart contract security.

The interview comprised three parts. In the first part, we asked some demographic questions about the experience of the interviewees in smart contract development. The questions covered various aspects of experience, including programming, design, testing, and project management.

In the second part, we asked open-ended questions about the security awareness and practices of smart contract development. The purpose of this part was to allow the interviewees to speak freely about their opinions and experience without the interviewer biasing their responses.

In the third part, we asked the interviewees to discuss the sources where they obtain security-related knowledge, as well as strategies and tools that they have used for security assurance of smart contracts in the practices.

At the end of each interview, we thanked the interviewee and briefly informed her of our next plans.

\subsubsection{Participant Selection}
We recruited full-time software practitioners with experience in smart contract development from blockchain companies (e.g., Hyperchain\footnote{\url{https://www.hyperchain.cn/en}}), IT companies (e.g., Alibaba) and open-source smart contract projects. 
Interviewees were recruited by emailing our contact in each company or project, who disseminated the news of our study to their colleagues. Volunteers would inform us if they were willing to participate in the study with no compensation. 
With this approach, 13 volunteers with varied experience in years contacted us -- 7 interviewees from four companies and 6 interviewees from three open-source projects. In the remainder of this paper, we denote these 13 interviewees as P1 to P13. These 13 interviewees have an average of 6.5 years of professional experience in software development (min: 3, max: 13, median: 6.5, sd: 2.7), and an average of 2.3 years in smart contract development (min: 1, max: 5, median: 2, sd: 1.1). Table \ref{tab:extensive_exp} summarizes the number of interviewees who perceived themselves as having  ``extensive'' experience (in comparison to ``none'' and ``some'' experience) in a particular role.

\subsubsection{Data Analysis}
We conducted a thematic analysis~\cite{cruzes2011recommended} to process the recorded interviews by following the steps below:

\noindent\textbf{Transcribing and Coding.} 
After the last interview was completed, we transcribed the recordings of the interviews, and developed a thorough understanding by reviewing the transcripts.
The first author read the transcripts and coded the interviews using NVivo qualitative analysis software \cite{qsr2020nvivo}.

To ensure the quality of codes, the second author verified initial codes created by the first author and provided suggestions for improvement.
After incorporating these suggestions, we generated a total of 427 cards that contain the codes - 30 to 41 cards for each coded interview.
After merging the codes with the same words or meanings, we have a total of 133 unique codes.

\noindent\textbf{Open Card Sorting.}
Two of the authors then separately analyzed the codes and sorted the generated cards into potential themes for thematic similarity (as illustrated in LaToza et al.'s study \cite{latoza2006maintaining}). The themes that emerged during the sorting were not chosen beforehand. We then use the Cohen's Kappa measure \cite{cohen1960coefficient} to examine the agreement between the two labelers.
The overall Kappa value between the two labelers is 0.76, which indicates substantial agreement between the labelers. After completing the labeling process, the two labelers discussed their disagreements to reach a common decision. To reduce bias from the two authors sorting the cards to form initial themes, they both reviewed and agreed on the final set of themes. Eventually, we derived 6 competing priorities, a list of 11 security motivators and 9 security deterrents, 5 sources of security knowledge, and 11 security strategies, and 11 factors that affect the adoption of security tools.  

\begin{table}[t]
  \centering
  \caption{Number of interviewees with ``extensive'' experience in a particular role.}
  
    \begin{tabular}{lcc}
        \Xhline{2\arrayrulewidth}
          \multicolumn{1}{l}{\textbf{Role}} & \multicolumn{1}{l}{\textbf{Smart Contract}} & \multicolumn{1}{l}{\textbf{non-Smart-Contract}}\\
    \hline
    Programming & 10     & 12 \\
    Design & 8     & 6 \\
    Management & 3     & 4 \\
    Testing & 3     & 3 \\
       \Xhline{2\arrayrulewidth}
    \end{tabular}\label{tab:extensive_exp}\vspace{-0.5cm}
\end{table}

\subsection{Survey}
\label{sec:survey}
The right part of Fig.~\ref{fig:process} describes the process of our online survey.
\subsubsection{Protocol}
We conducted an IRB-approved anonymous online survey with professional smart contract practitioners. The survey aims to validate and quantify the observations from our interviews.      
We followed Kitchenham and Pfleeger's guidelines for personal opinion surveys \cite{kitchenham2008POS} and used an anonymous survey to increase response rates \cite{tyagi1989TEO}. A respondent has the option to specify that she prefers not to answer or does not understand the description of a particular question. We include this option to reduce the possibility of respondents providing arbitrary answers.

\noindent\textbf{Recruitment of Respondents.} 
To recruit respondents from the population of smart contract practitioners, we spread the survey to a broad range of companies from various locations worldwide. No identifying information was required or gathered from our respondents.
To get a sufficient number of respondents from diverse backgrounds, we followed a multi-pronged strategy to recruit respondents:
\begin{itemize}
\item We contacted professionals from blockchain companies and IT companies that launch blockchain projects around the world and asked their help to disseminate our survey within their organizations. 
Specifically, we sent emails to our contacts in Alibaba, Baidu, Hengtian, Hyperchain, IBM, Morgan Stanley, and other companies, encouraging them to disseminate our survey to some of their colleagues who have experience in smart contract development. 
By following this strategy, we aimed to recruit respondents working with smart contracts in the industry from diverse organizations.

\item We sent an email with a link to the survey to 1,986 practitioners who contributed to 12 blockchain repositories that support smart contracts (e.g., {\emph{ethereum/go-ethereum}, \emph{EOSIO/eos} and \emph{hyperledger/fabric}}) and 580 smart contract repositories (e.g., \emph{ethereum/solidity} and \emph{EOSIO/eosio.contracts}) hosted on GitHub and solicited their participation. We aimed to recruit open-source practitioners who have smart contract experience in addition to professionals working in the industry.

Out of these emails, eight emails received automatic replies notifying us of the absence of the receiver; two emails indicated the receivers left the original organizations; four receivers replied that they only have experience in blockchain but not smart contract development.
\end{itemize}

\subsubsection{Survey Design}
The survey includes different types of questions, e.g., multiple-choice and free-text answer questions. The potential answers and statements of multiple-choice questions were derived from the results of our interviews. For these questions, we include an ``\emph{I don't know}'' option in case some statements are not applicable to the experience of respondents, or respondents had a poor understanding of the statements. 

The survey consists of four sections, grouping questions by topic to minimize the cognitive load on participants and allow them to consider the topic more deeply \cite{lazar2010research}. 
Specifically, the following four sections have been captured in the survey (the complete questionnaire is available online as supplemental material\footnote{Questionnaire  Online: \url{http://doi.org/10.5281/zenodo.4005112}}):

\noindent\textbf{Demographics.}
We collected demographic information about the respondents to allow us to (1) filter respondents who may not understand our survey (i.e., respondents without any experience in smart contracts), (2) breakdown the results by groups (e.g., public, consortium, and private blockchains).
Specifically, we asked two questions:
\begin{itemize}
  \item \emph{Do you have experience with smart contracts?}
  \item \emph{What best describes the \textbf{primary blockchain platform} that you currently work on}?
\end{itemize}
In terms the second question, we provided four options for primary blockchain platforms, including (1) \emph{public blockchain}, (2) \emph{consortium blockchain}, (3) \emph{private blockchain}, and (4) \emph{other}. 

Based on the selections of respondents, we could exclude invalid responses and divide the survey respondents into three groups.
To focus the respondents' attention on a particular blockchain platform in the survey, they were explicitly asked to answer each following question with respect to their experience with the \emph{primary blockchain platform} they specified.

\begin{figure}
\center
\includegraphics[scale=0.55]{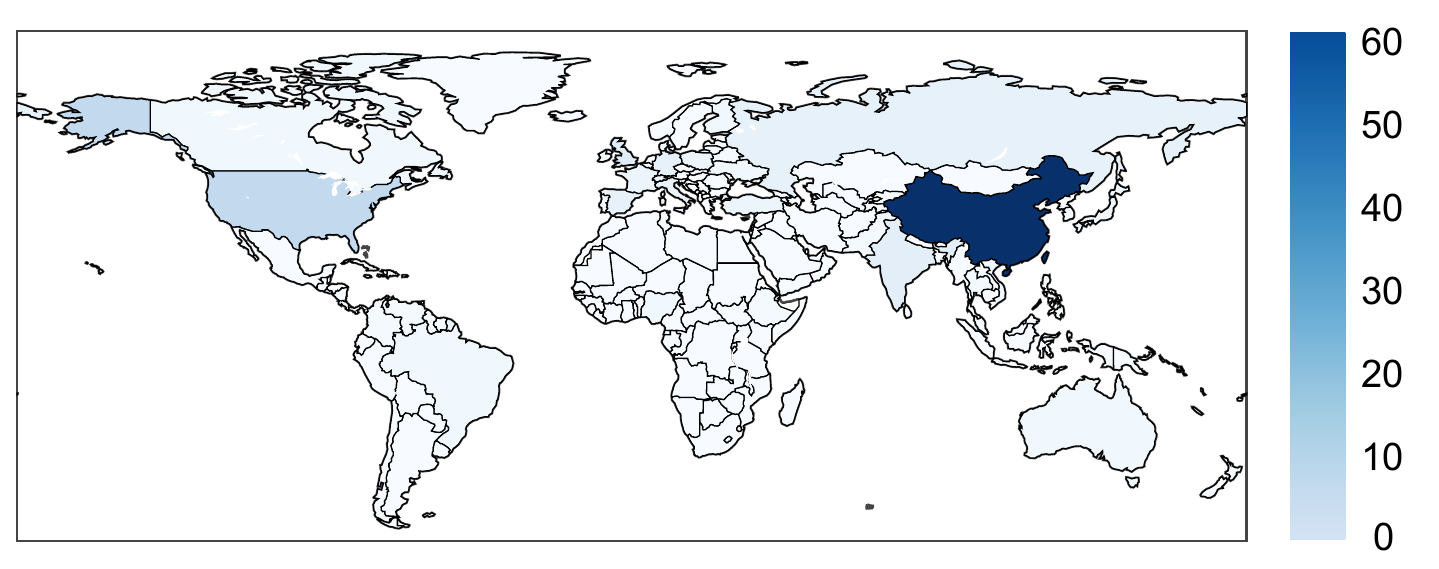}\vspace{-0.2cm}
\caption{Countries in which survey respondents reside. The darker the color is, the more respondents reside in that country. The legend indicates the number of respondents.}
\label{fig:map}\vspace{-0.5cm}
\end{figure}

\begin{figure}
  \center
  \includegraphics[scale=0.37]{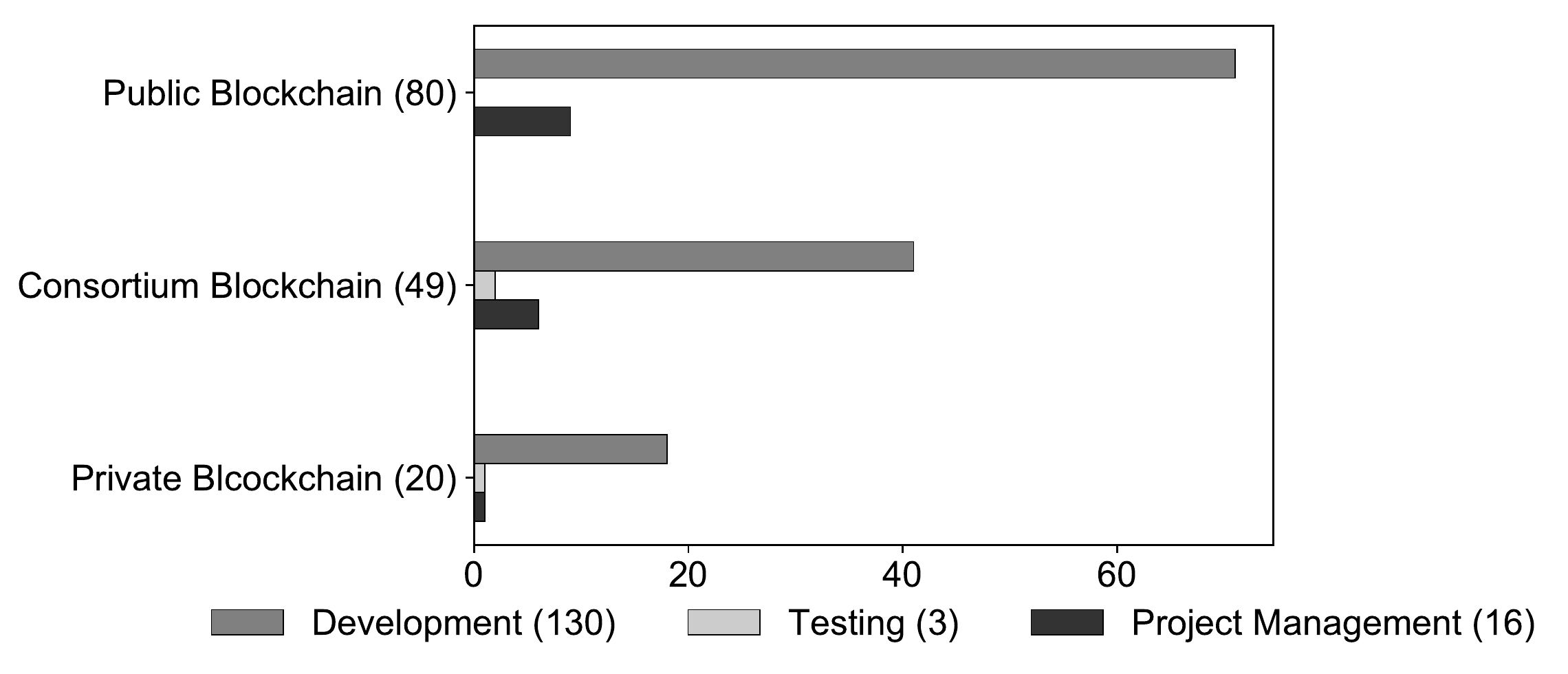}\vspace{-0.5cm}
  \caption{Survey respondents demographics. The number indicates the count of each demographic group.}
  \label{fig:job}\vspace{-0.5cm}
  \end{figure}

We received a total of 203 responses, and further excluded 46 responses made by respondents who claimed that they do not have experience in smart contract development. We also excluded one response made by a respondent who described her job role as sales. In the end, we had a set of 156 valid responses.
The 156 respondents reside in 35 countries across six continents as shown in Fig.~\ref{fig:map}. The top two countries in which the respondents reside are China (61) and the United States (16). The respondents have an average of 6.3 years of professional experience (min: 0.5, max: 40, median: 4, sd: 6.9), with an average of 2 years of experience in smart contract development (min: 0.1, max: 6, median: 2, sd: 1.4).
Our survey respondents are distributed across different demographic groups (job roles and primary blockchain platforms) as shown in Fig.~\ref{fig:job}. Seven respondents who selected \emph{Other} as their primary blockchain platforms and explained that they simultaneously work on more than one blockchain. We excluded the responses of the seven respondents from any comparisons between groups of different blockchains.

\noindent\textbf{Perceptions on Smart Contract Security.} This section investigates practitioners' perceptions of smart contract security, specifically, the importance of security, awareness of security problems, as well as the motivators and deterrents to smart contract security.

\noindent\textbf{Security Practices in Smart Contract Development.} This section focused on security practices in smart contracts, including practitioners' efforts towards security, their strategies for achieving security, and tools for securing smart contracts. 

More details about the questions and format are available in Section \ref{sec:result}, along with the corresponding results.

We piloted the preliminary survey with a small set of smart contract practitioners who were different from our interviewees and survey takers. We obtained feedback on (1) whether the length of the survey was appropriate, and (2) the clarity and understandability of the terms. We made minor modifications to the preliminary survey based on the received feedback and produced a final version. Note that the collected responses from the pilot survey are excluded from the presented results in this paper.

To support respondents from China, we translated our survey to Chinese before publishing the survey. We chose to make our survey available both in English on Google Forms, and in Chinese on a popular survey website in China\footnote{https://www.wjx.cn}. The reason is that English is an international lingua franca, and Chinese is the most spoken language. We expected that a large number of our survey recipients are fluent in one of these two languages. We carefully translated our survey to make sure there exists no ambiguity between English and Chinese terms in our survey. Also, we polished the translation by improving clarity and understandability according to the feedback from our pilot survey.

\subsubsection{Data Analysis} We analyzed the survey results based on the question types. For multiple-choice questions, we reported the percentage each option is selected. 
In terms of open-ended questions, we followed an inductive approach in which two authors separately performed open card sorting and regularly discussed emerging themes until an agreement was reached.

\noindent\textbf{Factor Analysis.} To identify meaningful clusters of closely related information, we used factor analysis to analyze the Likert-scale ratings of the statements with respect to the security motivators and security deterrents in smart contract development. Specifically, we used principal axis factor analysis in the {\tt psych} R library~\cite{psych2017} to group related information with a cut-off point of $|0.4|$ for factor loading. We used the {\tt fa.parallel} function in the {\tt psych} R library to select the optimal number of factors for factor analysis. Thus, we reduced a large set of variables into a smaller set (\emph{factors}) while retaining the majority of original information~\cite{thompson2007exploratory}.

\noindent\textbf{Comparison.} We classified our respondents into different groups based on their primary blockchain platforms (i.e., public, consortium, and private blockchains), and compared the survey results of different groups of respondents. For instance, we used the Wilcoxon rank-sum test for Likert-scale answers to perform the comparison. 
All statistical tests assumed a p-value $< 0.05$ as a significant level. Bonferroni correction was applied to adjust p-values in multiple comparisons. 
\section{Results}\label{sec:result}
We explain the results of three research questions that investigate smart contract security from the perspective of practitioners.
\subsection{RQ1: Perceptions of Smart Contract Security}
In RQ1, we explored practitioners' priorities in smart contract development, what motivates them and deters them to address smart contract security, and their experience of security problems. 
To understand practitioners' priorities in smart contract development, we presented our respondents with six statements that describe the requirements of smart contracts. Respondents ranked the importance of each requirement on a 5-point Likert scale (\emph{not at all}, \emph{slightly}, \emph{moderately}, \emph{very}, \emph{extremely}). 
To explore what drives practitioners to address smart contract security, we presented our respondents with a list of 11 statements that describe potential security motivators and 9 statements that explain potential security deterrents. Respondents indicated their level of agreement with each statement on a 5-point Likert scale (\emph{strongly disagree}, \emph{disagree}, \emph{neutral}, \emph{agree}, \emph{strongly agree}). 
In addition, we asked the respondents to report whether their smart contracts have ever experienced a security problem as well as the sources where they gain security knowledge.

\begin{figure}[t]
  \center
  \includegraphics[scale=0.52]{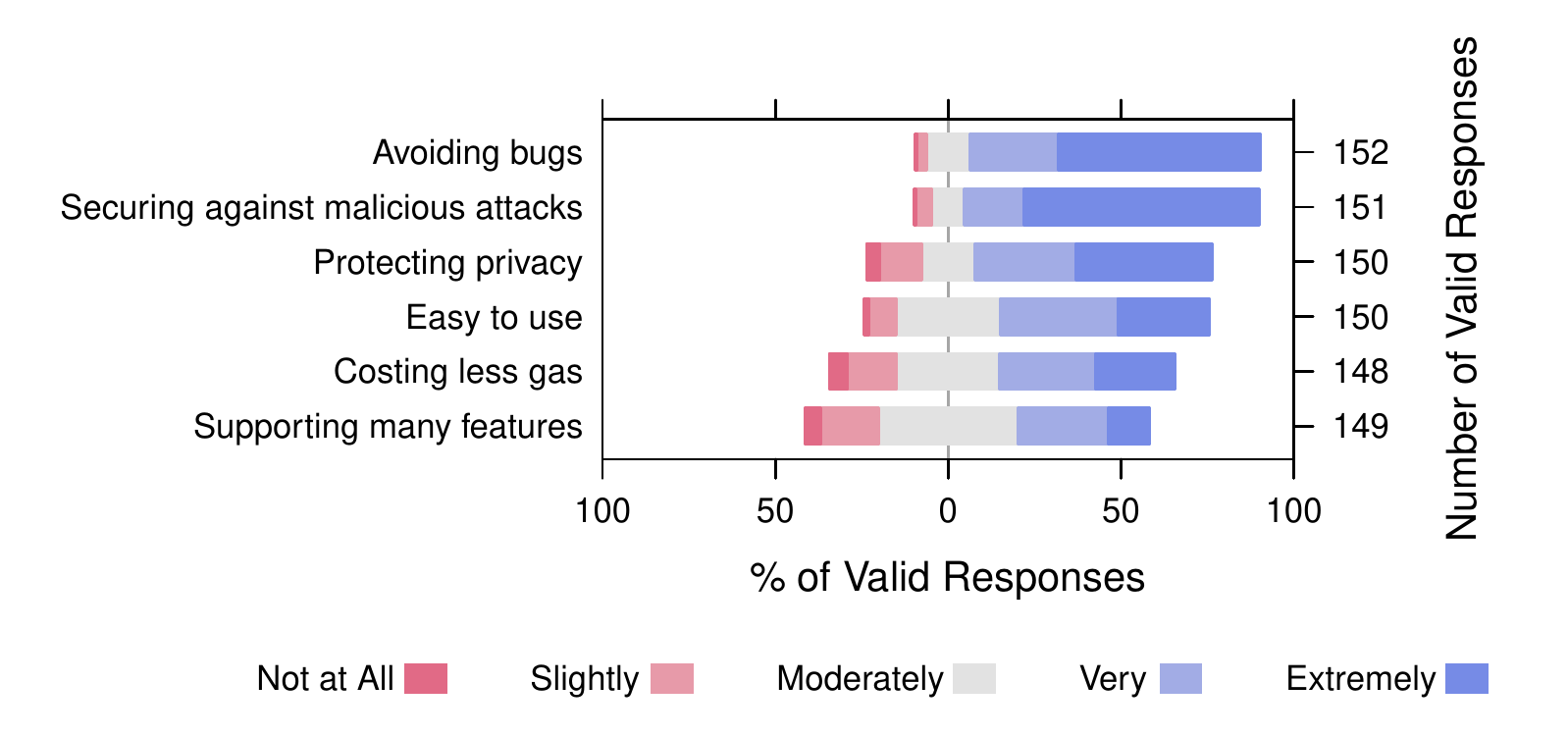}\vspace{-0.2cm}
  \caption{Importance of Different Requirements.}
  \label{fig:importance}\vspace{-0.5cm}

\end{figure}

\begin{figure*}[htbp]
  \center
  \hspace{0.2in}\includegraphics[scale=0.5]{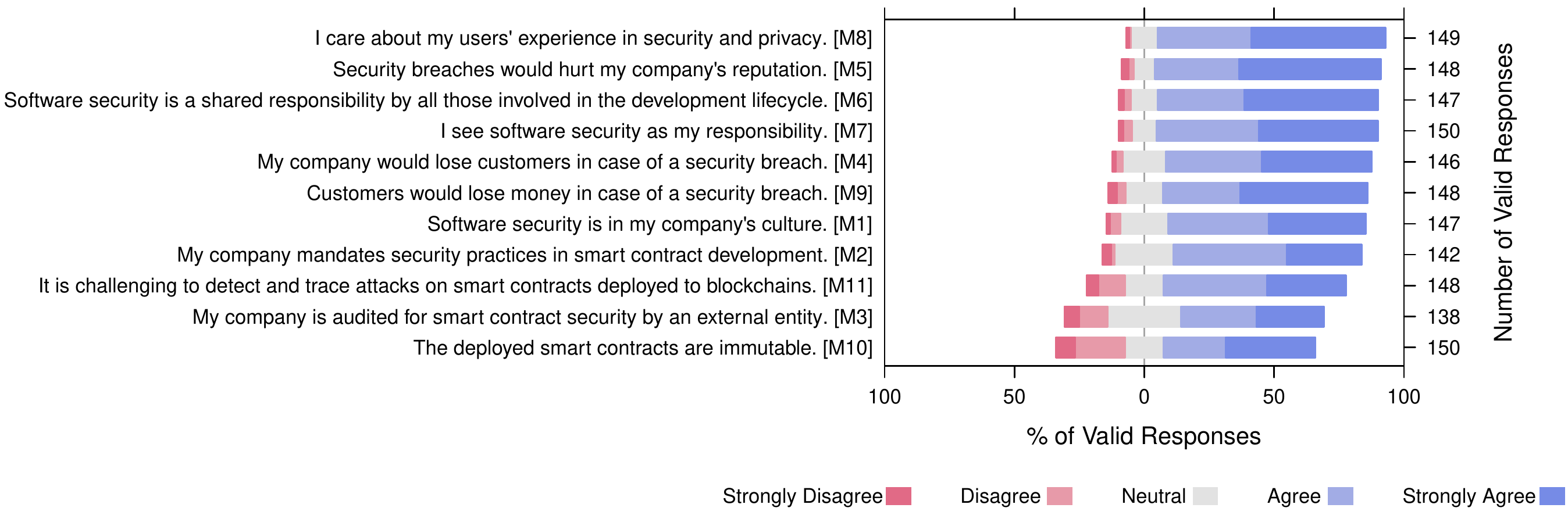}\vspace{-0.2cm}
  \caption{Motivators of smart contract security.}
  \label{fig:motivators}\vspace{-0.5cm}

\end{figure*}

\noindent\textbf{Importance of Security.} Fig.~\ref{fig:importance} shows respondents' ratings of the importance of various requirements in their smart contracts. In addition to avoiding bugs, 85\% and 69\% of the respondents considered security and privacy \emph{very} or \emph{extremely} important, respectively. The ratings were higher than the requirement of costing less gas in smart contracts.

\noindent\textbf{Security Motivators.} We asked the respondents ``\emph{I care about smart contract security because ...}'' and presented the 11 potential motivators for smart contract security. 
As shown in Fig.~\ref{fig:motivators}, the top two security motivators are respondents' internal motivations\footnote{Internal motivation: people stand behind a behavior out of their interests and values~\cite{ryan2000self}.}, i.e., to protect their users and the reputation of their companies. Meanwhile, external motivations\footnote{External motivation: people do a behavior for reasons external to the self~\cite{ryan2000self}.} are reportedly less motivating, i.e., the immutability of smart contracts and external auditing. 

We used factor analysis to cluster the 11 motivators into three factors as shown in Table~\ref{tab:motivator}. Two motivators did not conform to any particular factor. We named the factors as \emph{awareness of importance}, \emph{workplace environment} and \emph{perceived negative consequences}. Out of the three factors, \emph{workplace environment} is the only external motivation.

\begin{figure*}[htbp]
    \center
    \includegraphics[scale=0.5]{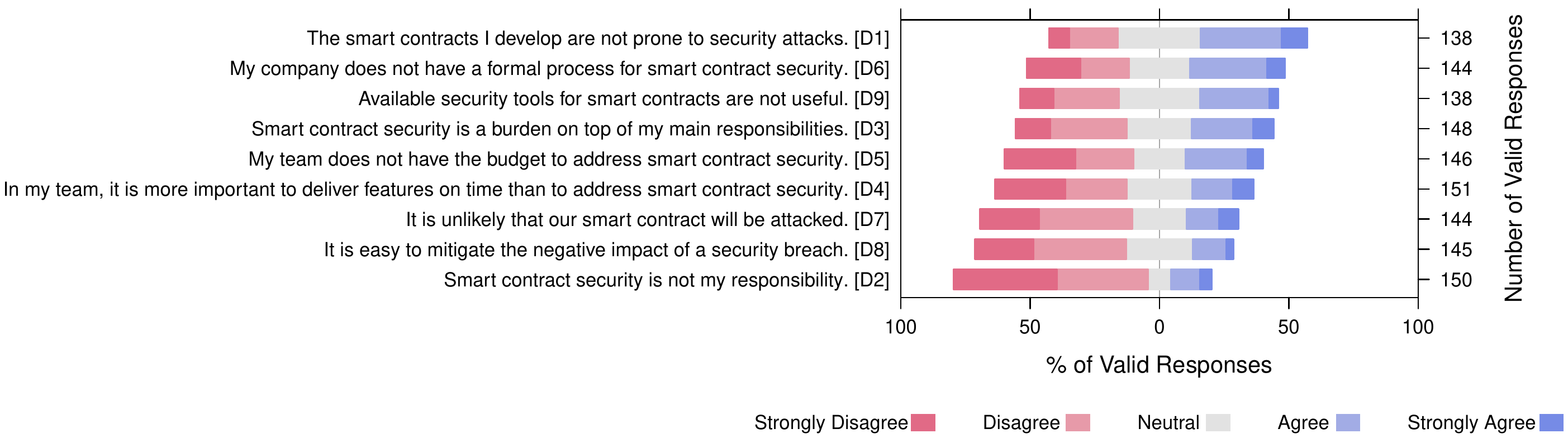}\vspace{-0.2cm}

    \caption{Deterrents of smart contract security.}

    \label{fig:deterrents}\vspace{-0.5cm}

    \end{figure*}

\begin{table}[t]
    \centering
    \caption{Factor analysis of motivators to smart contract security.}
      \begin{tabular}{D C}
        \Xhline{2\arrayrulewidth}
        
        \textbf{Variables (Motivators as presented in the survey)} & \textbf{Factor loading} \\
        \hline
        \rowcolor[rgb]{ .851,  .851,  .851}\multicolumn{2}{c}{\textit{\textbf{Awareness of Importance}}} \\
        \hline
        {[M7]} I see software security as my responsibility. & 0.76 \\
        {[M6]} Software security is a shared responsibility by all those involved in the development lifecycle. & 0.71 \\
        {[M8]} I care about my users' experience in security and privacy. & 0.68 \\
        {[M1]} Software security is in my company's culture. & 0.51 \\
        \hline
        \rowcolor[rgb]{ .851,  .851,  .851}\multicolumn{2}{c}{\textit{\textbf{Workplace Environment}}} \\
        \hline
        {[M3]} My company is audited for smart contract security by an external entity. & 0.73 \\
        {[M2]} My company mandates security practices in smart contract development. & 0.61 \\
        \hline
        \rowcolor[rgb]{ .851,  .851,  .851}\multicolumn{2}{c}{\textit{\textbf{Perceived Negative Consequences}}} \\
        \hline
        {[M4]} My company would lose customers in case of a security breach. & 0.78 \\
        {[M5]} Security breaches would hurt my company's reputation. & 0.66 \\
        {[M9]} Customers would lose money in case of a security breach. & 0.56 \\
        \hline
        \rowcolor[rgb]{ .851,  .851,  .851}\multicolumn{2}{c}{\textit{\textbf{Motivators not belonging to any factor}}} \\
        \hline
        {[M10]} The deployed smart contracts are immutable. &  \\
        {[M11]} It is challenging to detect and trace attacks on smart contracts deployed to blockchains. &  \\
      \Xhline{2\arrayrulewidth}
      \end{tabular}\label{tab:motivator} \vspace{-0.3cm}
  \end{table}

\noindent\textbf{Security Deterrents.} Respondents generally opposed statements that imply deferring or ignoring security, as suggested by the longer red bars in comparison with blue bars (Fig.~\ref{fig:deterrents}). The top two deterrents of smart contract security are a lack of awareness of security attacks, followed by a formal process.

Our factor analysis combined 8 out of the 9 deterrents into two factors; 1 deterrent did not correspond to any particular factor (Table~\ref{tab:deterrent}). The first factor \emph{competing priorities and no process} describes how a lack of security can arise from systemic causes within an organization or a team. The other factor \emph{irrelevance of security} characterizes the personal-level awareness of security risks that can deter practitioners from addressing smart contract security. 

\begin{table}[t]
    \centering
    \caption{Factor analysis of deterrents to smart contract security.}
    \begin{tabular}{D C}
        \Xhline{2\arrayrulewidth}
      \textbf{Variables (Deterrents as presented in the survey)} & \textbf{Factor loading} \\
      \hline
      \rowcolor[rgb]{ .851,  .851,  .851}\multicolumn{2}{c}{\textit{\textbf{Competing Priorities and No Process}}} \\
      \hline
      {[D5]} My team does not have the budget to address smart contract security. & 0.86 \\
      {[D6]} My company does not have a formal process for smart contract security. & 0.79 \\
      {[D4]} In my team, it is more important to deliver features on time than to address smart contract security. & 0.74 \\
      {[D2]} Smart contract security is not my responsibility. & 0.52 \\
      {[D3]} Smart contract security is a burden on top of my main responsibilities. & 0.43 \\
      \hline
      \rowcolor[rgb]{ .851,  .851,  .851}\multicolumn{2}{c}{\textit{\textbf{Irrelevance of Security}}} \\
      \hline
      {[D7]} It is unlikely that our smart contract will be attacked. & 0.64 \\
      {[D1]} The smart contracts I develop are not prone to security attacks. & 0.58 \\
      {[D8]} It is easy to mitigate the negative impact of a security breach. & 0.56 \\
      \hline
      \rowcolor[rgb]{ .851,  .851,  .851}\multicolumn{2}{c}{\textbf{Deterrents not belonging to any factor}} \\
      \hline
      {[D9]} Available security tools for smart contracts are not useful. &  \\
      \Xhline{2\arrayrulewidth}
      \end{tabular}\label{tab:deterrent}\vspace{-0.5cm}
  \end{table}

\noindent\textbf{Experiencing Security Problems.} 40\% of our respondents reported that they had experienced at least one out of three potential security problems, i.e., vulnerabilities in unshipped code, vulnerabilities in shipped code, and security breaches. Identification of vulnerable code before smart contracts were shipped was the most frequently reported (22\%) security problem in our survey. 19\% of the respondents indicated that vulnerabilities were discovered in shipped smart contracts. 10\% reported that their smart contracts experienced a security breach. We note that these numbers are not mutually exclusive; 10\% of the respondents reported multiple security problems.

\noindent\textbf{Sources of Security Knowledge.} Official forums of blockchain platforms (60\%), research papers (53\%), question and answer websites (47\%) are the top three most popular sources for respondents to acquire security knowledge about smart contracts. We note that these numbers are not mutually exclusive; 74\% of our respondents use more than one source to gain knowledge of smart contract security.

\subsection{RQ2: Security Practices in Smart Contract Development}

The survey had several questions exploring the efforts and strategies that development teams employ to ensure smart contract security. 44\% of our respondents received support from professional security experts.

\noindent\textbf{Efforts towards Security.} Respondents reported the percentage of efforts directed towards security out of the overall efforts in the development lifecycle of smart contracts. They also reported to what extent security was considered for each stage in the development lifecycle (i.e., requirement, design, construction, testing, deployment, and maintenance).

Our respondents indicated that, on average, security efforts account for 29\% (min: 0\%, max: 100\%, median: 20\%, sd: 26\%) of the overall efforts in smart contract development. 14 respondents indicated that their teams do not spend any effort on security. 

\begin{figure}[t]
    \hspace{0.58cm}\includegraphics[scale=0.55]{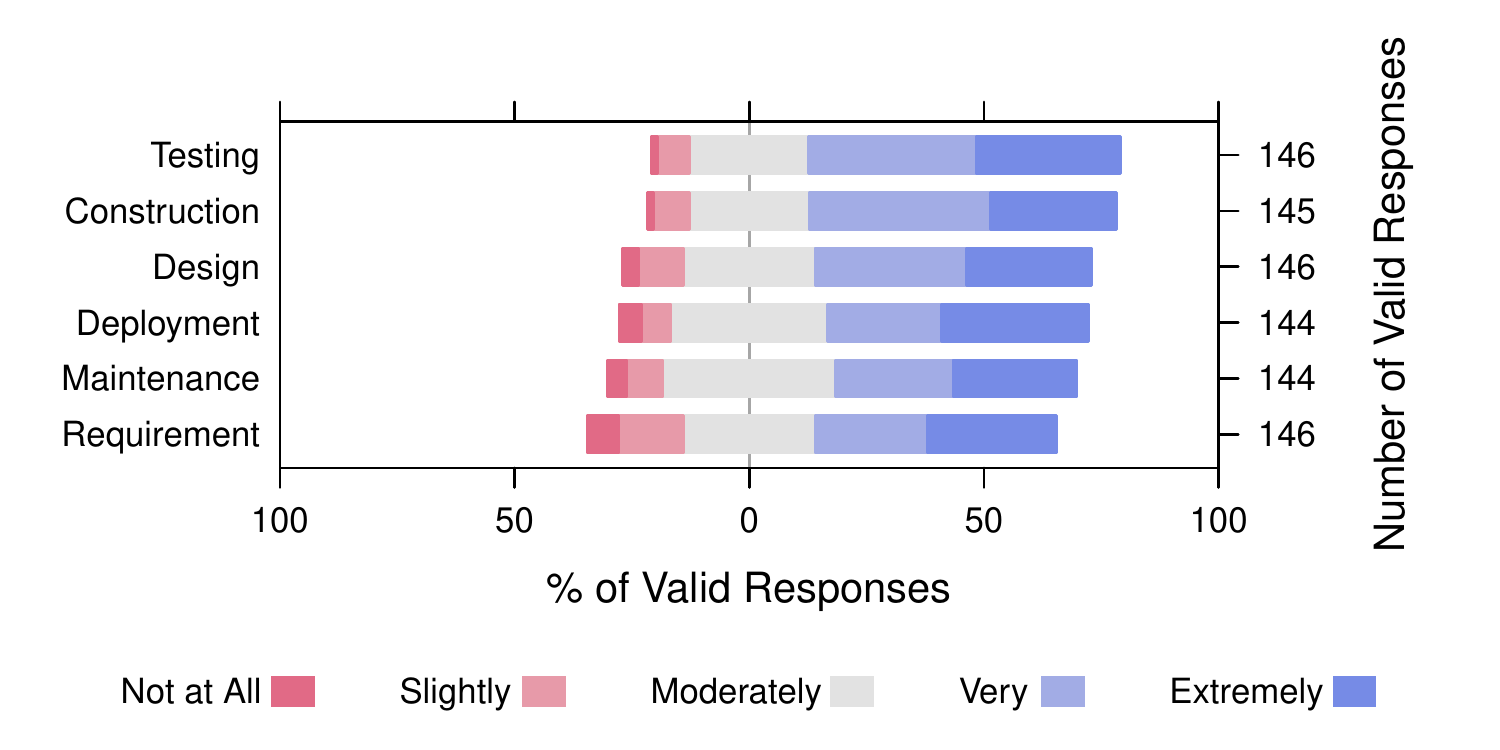}\vspace{-0.2cm}
    \caption{Security efforts across stages in development lifecycle.}
    \label{fig:stage}\vspace{-0.5cm}
\end{figure}

We used the Wilcoxon rank sum test to determine if the distribution of security efforts statistically significantly differs across different stages in the process.  As shown in Fig.~\ref{fig:stage},
security effort at the testing stage was statistically significantly higher than that at the requirement (p-value = 0.01) and maintenance stages (p-value = 0.04).
Security effort at the construction stage was statistically significantly higher than that at the requirement stage (p-value = 0.05).  
Although our interviewees (P4 and P6) mentioned that ``\emph{we try to get it (security) right from the beginning}'', security effort at the requirement stage was ranked at the bottom across different stages in the development lifecycle.

\noindent\textbf{Strategies to Address Smart Contract Security.} 
We provided a list of 11 statements that describe potential security strategies, and asked the respondents to rate each statement from the list on a 5-point Likert scale (\emph{very rarely}, \emph{rarely}, \emph{sometimes}, \emph{often}, \emph{very often}).

\begin{figure}[t]
    \includegraphics[scale=0.52]{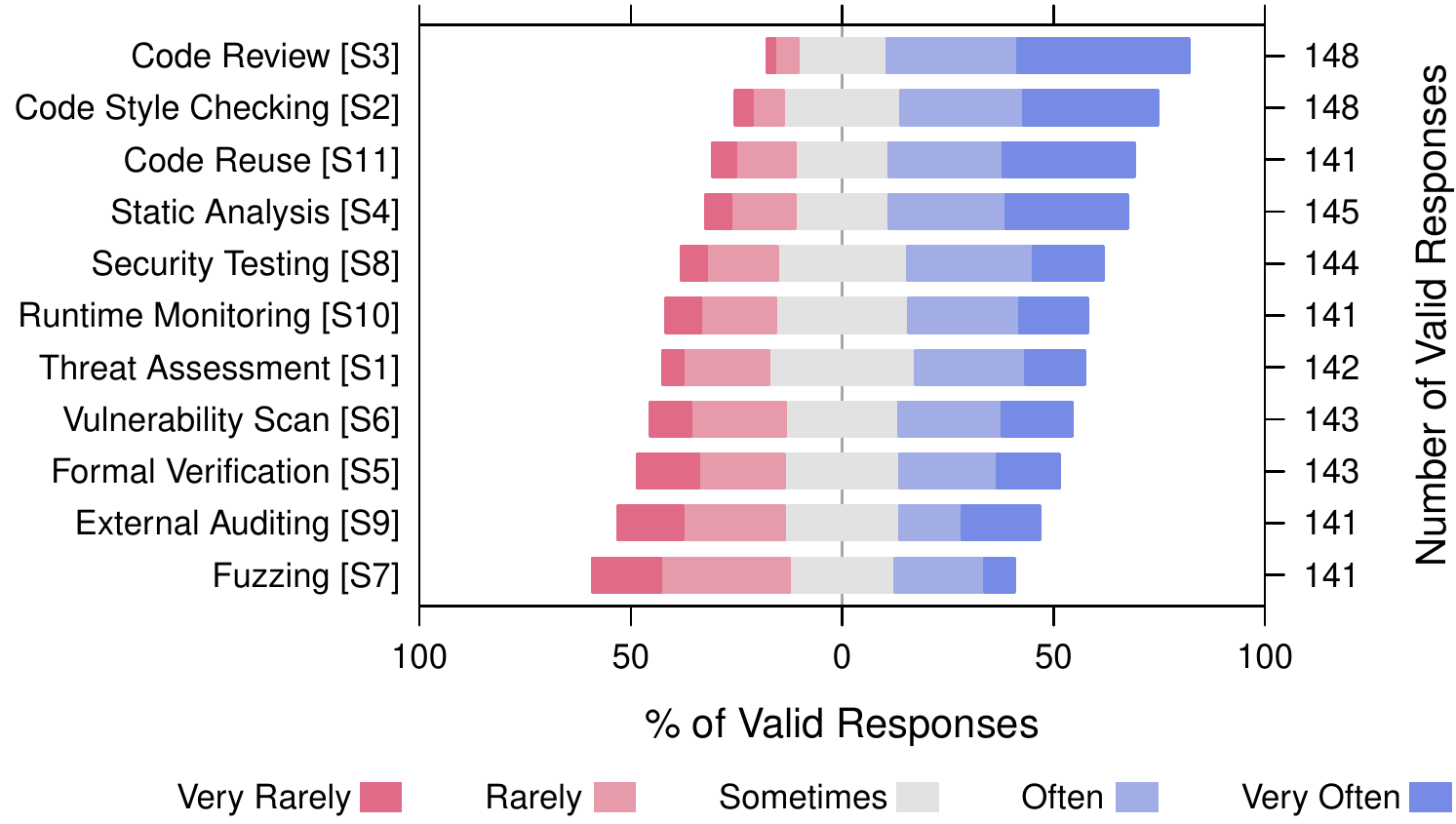}\vspace{-0.2cm}
    \caption{Strategies for handling smart contract security.}
    
    \label{fig:strategy}\vspace{-0.6cm}

\end{figure}

Our respondents combine various strategies to address smart contract security in practice. 72\% of our respondents frequently leverage more than one security strategy in smart contract development. As shown in Fig.~\ref{fig:strategy}, code review is the most frequently used security strategy  -- 72\% of our respondents indicated that they \emph{often} or \emph{very often} rely on code review to address smart contract security. 61\% and 58\% of the respondents \emph{often} or \emph{very often} do code style checking and reuse code from reliable sources, respectively. Only 28\% of the respondents \emph{often} or \emph{very often} integrate fuzzing into the development lifecycle.

A total of 24 respondents provided free-form text comments regarding other security strategies they use in practice. Out of the 24 respondents, 11 drilled down the aforementioned strategies; the other 13 respondents identified additional strategies (followed by their corresponding frequency) as follows:
\begin{itemize}
    \item Security by Design (5): ``\emph{Security concerns should be built into the framework and exposed via documented, developer-friendly APIs so that good security is easy and bad security is hard.}''
    \item Programming Languages (3): ``\emph{... use the most stable version of Solidity avoiding the latest one.}''
    \item Dependency Management (2): ``\emph{Dependency management to ensure we're using recent versions.}''
    \item Learning from Past Experiences (2) ``\emph{... failure code of others in the past.}''
    \item Seeking Support from Experts (1): ``\emph{... ensured by cryptography designing together with experts.}''
    
\end{itemize}

\noindent\textbf{Tools to Address Smart Contract Security.} We further investigated the adoption of security tools for smart contracts. 54\% of our respondents indicated that they frequently adopt security tools in smart contract development. Security plugins in Integrated Development Environments (IDE) are the most popular security tool for smart contracts -- 45\% of our respondents indicated that they \emph{often} or \emph{very often} rely on security plugins in IDEs to address smart contract security. We further investigated how frequently the security tools for smart contracts have been adopted in practice. 19\%, 12\%, 14\% and 12\% of the respondents reported that they \emph{often} or \emph{very often} use \emph{Mythril}~\cite{consensys2019Mythril}, \emph{Oyente}~\cite{luu2016making}, \emph{SmartCheck}~\cite{tikhomirov2018smartcheck} and \emph{Slither}~\cite{feist2019slither}, respectively. 

\begin{figure}[t]
    \center
    \includegraphics[scale=0.52]{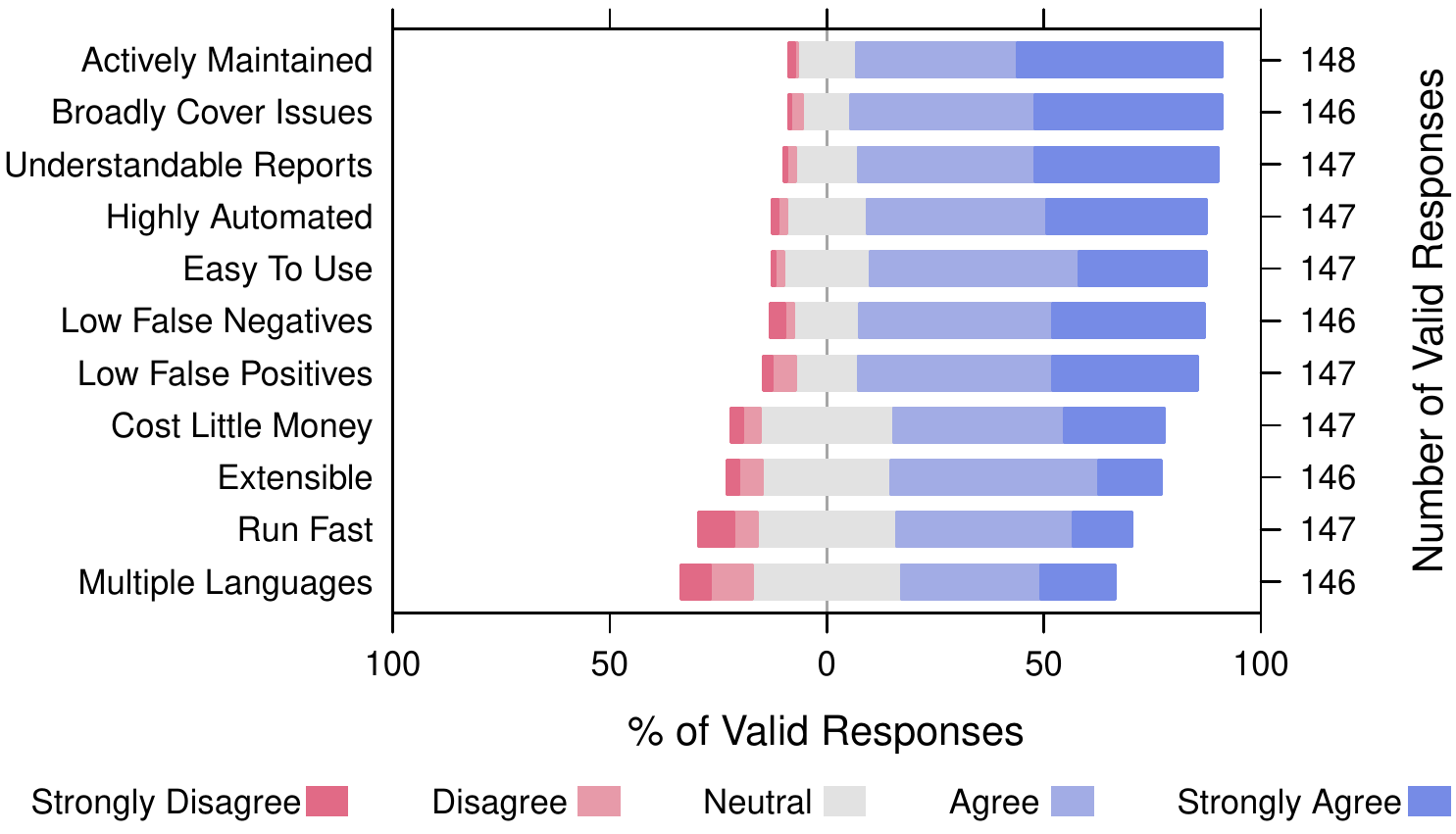}\vspace{-0.2cm}
    \caption{Factors that affect adoption of security tools.}
    \label{fig:tool_adoption}\vspace{-0.6cm}

\end{figure}

In addition, we investigated what factors affect the adoption of security tools in smart contract development. We provided a list of 11 statements that describe potential factors and asked the respondents to rate each statement from the list on a 5-point Likert scale (\emph{strongly disagree}, \emph{disagree}, \emph{neutral}, \emph{agree}, \emph{strongly agree}). As shown in Fig.~\ref{fig:tool_adoption}, active maintenance is the most important factor in the adoption of security tools (85\% \emph{agree} or \emph{strongly agree}). 86\% of the respondents \emph{agree} or \emph{strongly agree} that coverage of security issues affects the adoption of security tools.

\begin{table*}[t]
  \centering
  \caption{Impact of blockchain platforms on smart contract security. \colorbox{cellorange}{Orange} cells indicate where the former group is more negative about the statement than the latter group; \colorbox{cellblue}{blue} cells indicate where the former group is more positive. \colorbox{cellgreen}{Green} cells represent statistically significant differences. The number in ``()'' indicates the size of each group.}
  \tiny
  \begin{tabular}{|r|l|ccc|ccc|}
      &       & \multicolumn{3}{c|}{\textbf{Effect Size}} & \multicolumn{3}{c|}{\textbf{P-value}} \\
      &       & \textbf{Public (80)} & \textbf{Public (80)} & \textbf{Consortium (49)} & \textbf{Public (80)} & \textbf{Public (80)} & \textbf{Consortium (49)} \\
      &       & \textbf{vs.} & \textbf{vs.} & \textbf{vs.} & \textbf{vs.} & \textbf{vs.} & \textbf{vs.} \\
      \textbf{Statement} &       & \textbf{Consortium (49)} & \textbf{Private (20)} & \textbf{Private (20)} & \textbf{Consortium (49)} & \textbf{Private (20)} & \textbf{Private (20)} \\
      \rowcolor[rgb]{ .851,  .851,  .851} \multicolumn{1}{|l|}{\textit{Motivators:}} &       &       &       &       &       &       &  \\
      Software security is in my company's culture.  & [M1]  & \cellcolor[rgb]{ .867,  .922,  .969}0.16 & 0.09  & -0.09 & 1.000 & 1.000 & 1.000 \\
      My company mandates security practices in smart contract development.  & [M2]  & \cellcolor[rgb]{ .867,  .922,  .969}0.16 & -0.02 & \cellcolor[rgb]{ 1,  .949,  .8}-0.20 & 1.000 & 1.000 & 1.000 \\
      My company is audited for smart contract security by an external entity.  & [M3]  & \cellcolor[rgb]{ .867,  .922,  .969}0.20 & 0.14  & -0.06 & 0.740 & 1.000 & 1.000 \\
      My company would lose customers in case of a security breach.  & [M4]  & \cellcolor[rgb]{ .867,  .922,  .969}0.18 & 0.09  & -0.11 & 0.778 & 1.000 & 1.000 \\
      Security breaches would hurt my company's reputation.  & [M5]  & 0.07  & 0.10  & 0.02  & 1.000 & 1.000 & 1.000 \\
      Software security is a shared responsibility by all those involved in the development lifecycle.  & [M6]  & -0.09 & -0.05 & 0.05  & 1.000 & 1.000 & 1.000 \\
      I see software security as my responsibility.  & [M7]  & \cellcolor[rgb]{ .867,  .922,  .969}0.18 & \cellcolor[rgb]{ .867,  .922,  .969}0.23 & 0.03  & 0.739 & 0.890 & 1.000 \\
      I care about my users' experience in security and privacy.  & [M8]  & \cellcolor[rgb]{ .867,  .922,  .969}0.16 & \cellcolor[rgb]{ .867,  .922,  .969}0.19 & 0.01  & 0.972 & 1.000 & 1.000 \\
      Customers would lose money in case of a security breach.  & [M9]  & 0.02  & -0.01 & -0.04 & 1.000 & 1.000 & 1.000 \\
      The deployed smart contracts are immutable.  & [M10] & \cellcolor[rgb]{ .608,  .761,  .902}0.39 & 0.14  & \cellcolor[rgb]{ 1,  .949,  .8}-0.24 & \cellcolor[rgb]{ .886,  .937,  .855}0.002 & 1.000 & 1.000 \\
      It is challenging to detect and trace attacks on smart contracts deployed to blockchains.  & [M11] & -0.13 & -0.10 & 0.01  & 1.000 & 1.000 & 1.000 \\
      \rowcolor[rgb]{ .851,  .851,  .851} \multicolumn{1}{|l|}{\textit{Deterrents:}} &       &       &       &       &       &       &  \\
      The smart contracts I develop are not prone to security attacks.  & [D1]  & \cellcolor[rgb]{ 1,  .949,  .8}-0.19 & -0.09 & 0.12  & 0.674 & 1.000 & 1.000 \\
      Smart contract security is not my responsibility.  & [D2]  & \cellcolor[rgb]{ 1,  .851,  .4}-0.36 & \cellcolor[rgb]{ 1,  .949,  .8}-0.28 & 0.08  & \cellcolor[rgb]{ .886,  .937,  .855}0.003 & 0.345 & 1.000 \\
      Smart contract security is a burden on top of my main responsibilities.  & [D3]  & 0.06  & \cellcolor[rgb]{ 1,  .949,  .8}-0.22 & \cellcolor[rgb]{ 1,  .949,  .8}-0.28 & 1.000 & 1.000 & 0.608 \\
      In my team, it is more important to deliver features on time than to address smart contract security.  & [D4]  & \cellcolor[rgb]{ 1,  .851,  .4}-0.43 & \cellcolor[rgb]{ .933,  .71,  0}-0.54 & -0.12 & \cellcolor[rgb]{ .886,  .937,  .855}0.000 & \cellcolor[rgb]{ .886,  .937,  .855}0.001 & 1.000 \\
      My team does not have the budget to address smart contract security.  & [D5]  & \cellcolor[rgb]{ 1,  .949,  .8}-0.23 & \cellcolor[rgb]{ 1,  .851,  .4}-0.35 & \cellcolor[rgb]{ 1,  .949,  .8}-0.21 & 0.268 & 0.130 & 1.000 \\
      My company does not have a formal process for smart contract security.  & [D6]  & \cellcolor[rgb]{ 1,  .949,  .8}-0.17 & \cellcolor[rgb]{ 1,  .949,  .8}-0.33 & \cellcolor[rgb]{ 1,  .949,  .8}-0.23 & 0.999 & 0.244 & 1.000 \\
      It is unlikely that our smart contract will be attacked.  & [D7]  & 0.03  & -0.12 & \cellcolor[rgb]{ 1,  .949,  .8}-0.16 & 1.000 & 1.000 & 1.000 \\
      It is easy to mitigate the negative impact of a security breach.  & [D8]  & -0.06 & \cellcolor[rgb]{ 1,  .949,  .8}-0.24 & \cellcolor[rgb]{ 1,  .949,  .8}-0.21 & 1.000 & 0.922 & 1.000 \\
      Available security tools for smart contracts are not useful.  & [D9]  & -0.02 & \cellcolor[rgb]{ 1,  .851,  .4}-0.34 & \cellcolor[rgb]{ 1,  .851,  .4}-0.37 & 1.000 & 0.280 & 0.223 \\
      \rowcolor[rgb]{ .851,  .851,  .851} \multicolumn{1}{|l|}{\textit{Security Efforts at Stages:}} &       &       &       &       &       &       &  \\
      Requirement & [E1]  & \cellcolor[rgb]{ .867,  .922,  .969}0.23 & 0.01  & \cellcolor[rgb]{ 1,  .949,  .8}-0.22 & 0.174 & 1.000 & 0.945 \\
      Design & [E2]  & \cellcolor[rgb]{ .867,  .922,  .969}0.27 & -0.06 & -0.34 & 0.059 & 1.000 & 0.186 \\
      Construction & [E3]  & \cellcolor[rgb]{ .867,  .922,  .969}0.28 & 0.00  & \cellcolor[rgb]{ 1,  .949,  .8}-0.25 & \cellcolor[rgb]{ .886,  .937,  .855}0.042 & 1.000 & 0.675 \\
      Testing & [E4]  & \cellcolor[rgb]{ .867,  .922,  .969}0.21 & \cellcolor[rgb]{ .867,  .922,  .969}0.22 & 0.09  & 0.255 & 0.806 & 1.000 \\
      Deployment & [E5]  & \cellcolor[rgb]{ .867,  .922,  .969}0.21 & -0.08 & \cellcolor[rgb]{ 1,  .949,  .8}-0.30 & 0.251 & 1.000 & 0.394 \\
      Maintenance & [E6]  & \cellcolor[rgb]{ .867,  .922,  .969}0.27 & 0.08  & \cellcolor[rgb]{ 1,  .949,  .8}-0.15 & 0.057 & 1.000 & 1.000 \\
      \rowcolor[rgb]{ .851,  .851,  .851} \multicolumn{1}{|l|}{\textit{Security Strategies:}} &       &       &       &       &       &       &  \\
      Threat Assessment & [S1]  & 0.06  & \cellcolor[rgb]{ .608,  .761,  .902}0.34 & \cellcolor[rgb]{ .867,  .922,  .969}0.32 & 1.000 & 0.265 & 0.435 \\
      Code Style Checking & [S2]  & 0.14  & \cellcolor[rgb]{ .867,  .922,  .969}0.17 & 0.03  & 1.000 & 1.000 & 1.000 \\
      Code Review & [S3]  & \cellcolor[rgb]{ .608,  .761,  .902}0.40 & \cellcolor[rgb]{ .608,  .761,  .902}0.38 & 0.09  & \cellcolor[rgb]{ .886,  .937,  .855}0.001 & 0.079 & 1.000 \\
      Static Analysis & [S4]  & 0.04  & \cellcolor[rgb]{ .867,  .922,  .969}0.26 & \cellcolor[rgb]{ .867,  .922,  .969}0.28 & 1.000 & 0.886 & 0.867 \\
      Formal Verification & [S5]  & -0.03 & 0.13  & \cellcolor[rgb]{ .867,  .922,  .969}0.19 & 1.000 & 1.000 & 1.000 \\
      Vulnerability Scan & [S6]  & -0.02 & \cellcolor[rgb]{ .867,  .922,  .969}0.17 & \cellcolor[rgb]{ .867,  .922,  .969}0.26 & 1.000 & 1.000 & 1.000 \\
      Fuzzing & [S7]  & -0.14 & 0.10  & \cellcolor[rgb]{ .867,  .922,  .969}0.24 & 1.000 & 1.000 & 1.000 \\
      Security Testing & [S8]  & 0.11  & \cellcolor[rgb]{ .867,  .922,  .969}0.24 & \cellcolor[rgb]{ .867,  .922,  .969}0.18 & 1.000 & 1.000 & 1.000 \\
      External Auditing & [S9]  & \cellcolor[rgb]{ .867,  .922,  .969}0.17 & \cellcolor[rgb]{ .867,  .922,  .969}0.19 & 0.04  & 1.000 & 1.000 & 1.000 \\
      Runtime Monitoring & [S10] & 0.06  & 0.13  & 0.10  & 1.000 & 1.000 & 1.000 \\
      Code Reuse & [S11] & \cellcolor[rgb]{ .867,  .922,  .969}0.28 & \cellcolor[rgb]{ .867,  .922,  .969}0.22 & -0.04 & 0.079 & 1.000 & 1.000 \\
\end{tabular}\label{tab:blockchain_comparison}\vspace{-0.3cm}\end{table*}

\subsection{RQ3: Effect of Blockchain Platforms on Security Perceptions and Practices}

In RQ3, We explore whether blockchain platforms influence security motivators and deterrents to smart contract security, as well as security efforts across different stages and strategies towards smart contract security. 

We summarize the results of comparisons in Table~\ref{tab:blockchain_comparison}. The \emph{Statement} column shows the statements presented to respondents in the survey. These statements describe the motivators, deterrents, stages in the development lifecycle, and security strategies. The following column indicates the labels we used to identify statements throughout the paper.

The \emph{Effect Size} column indicates the difference between \emph{Public Blockchain} and \emph{Consortium Blockchain} in the first subcolumn,  \emph{Public Blockchain} and \emph{Private Blockchain} in the second subcolumn, and \emph{Consortium Blockchain} and \emph{Private Blockchain} in the third subcolumn. We use Cliff's delta to measure the magnitude of the differences because Cliff's delta is reported to be more robust and reliable than Cohen's delta~\cite{romano2006appropriate}. Cliff's delta represents the degree of overlap between two sample distributions, ranging from $-1$ to $+1$. The extreme value $\pm 1$ occurs when the intersection between both groups is an empty set. When the compared groups tend to overlap, Cliff's Delta approaches zero. The magnitudes can be assessed with the thresholds as specified in~\cite{romano2006appropriate}: if $|\delta| < 0.147$, the effect size is negligible; if $0.147 \le |\delta| < 0.33$, the effect size is small; if $0.330 \le |\delta| < 0.474$, the effect size is medium; and otherwise the effect size is large. Effect sizes are additionally colored on a gradient from blue to orange based on the magnitudes of difference: blue color means the former group is more positive about the statement, and orange color means the latter group is more positive about the statement.

The \emph{P-value} column indicates whether the differences for each statement are statistically significant between \emph{Public Blockchain} and \emph{Consortium Blockchain} in the first subcolumn,  \emph{Public Blockchain} and \emph{Private Blockchain} in the second subcolumn, and \emph{Consortium Blockchain} and \emph{Private Blockchain} in the third subcolumn. Statistically significant differences at a 95\% confidence level (Bonferroni corrected p-value $< 0.05$) are highlighted in green.

Based on the observed statistically significant differences and effect  sizes, we can say with some certainty that:
\begin{itemize}
    \item \textbf{Security Motivators:} The blockchain platforms significantly impact the security motivator in terms of the immutability of smart contracts. The immutability of smart contracts drives practitioners of public blockchains more intensively than practitioners of consortium blockchains. In addition, the practitioners of public blockchains tend to be more motivated to address smart contract security than those of consortium and private blockchains.
    \item \textbf{Security Deterrents:} The blockchain platforms significantly affect the deterrents to security with respect to competing priorities in smart contract development. Practitioners of public blockchains tend to be more willing to prioritize security tasks over feature delivery and take the responsibility of addressing smart contract security, in comparison with those of consortium and private blockchains.
    
    \item \textbf{Security Efforts across Stages:} The blockchain platforms statistically significantly impact the security efforts at the construction stage in the development lifecycle. Practitioners of public blockchains spend more efforts towards security throughout the six stages in the development lifecycle, especially at the construction stage, in comparison with practitioners of consortium blockchains. In addition, practitioners of consortium blockchains tend to put less emphasis on security at the requirement, construction, deployment, and maintenance stages, in comparison with practitioners of private blockchains.
   
    \item \textbf{Security Strategies:} The blockchain platforms significantly affect the code review strategy that practitioners use to address smart contract security. Practitioners of public blockchains tend to perform code review more frequently than practitioners of consortium and private blockchains. Aside from code review, we observed no significant difference in the frequency of use of other security strategies between public and consortium blockchain practitioners. Private blockchain practitioners tend to use security strategies less frequently than public and consortium blockchain practitioners.
\end{itemize}

 \section{Discussion}\label{sec:discussion}
We reflect on our findings of research questions, delving into security awareness and risks of smart contracts, as well as code reuse and tool implications in smart contracts. We also highlight the avenues of future research across blockchain platforms.
\subsection{Security Awareness and Risks of Smart Contracts}
The vast majority of our respondents acknowledge the importance of smart contract security (RQ1). They prioritize security over the reduction of execution cost (e.g., gas consumption) in smart contract development. Our respondents spend 29\% of overall efforts on average in conducting security-related tasks. Previous studies found that developers generally exhibit a ``security is not my responsibility'' attitude~
\cite{xiao2014social}. On the contrary, 85\% of our respondents see smart contract security as their responsibility. \textbf{Smart contract practitioners tend to have a higher awareness of security than practitioners in other software areas.}

Despite the high awareness of security among smart contract practitioners, 40\% of our survey respondents indicated that their smart contracts suffered from security problems, including security breaches (RQ1). The percentage is higher than that of software in general (33\%) as reported in a recent survey~\cite{assal2019think}. The frequent occurrence of security problems in smart contracts may stem from the optimism bias~\cite{rhee2012unrealistic} ([D1]) and lack of a formal security process ([D6]) as suggested in RQ1. In addition, smart contracts on public blockchains are visible and accessible to all users, even malicious attackers. The inherent features of smart contracts make them more prone to security attacks than traditional software. Future work may focus on \textbf{standardizing and operationalizing the process of building security in smart contracts.}

\subsection{Code Reuse and Tool Implications in Smart Contracts}
58\% of the respondents frequently reuse code from reliable sources in smart contract development (RQ2). For instance, OpenZeppelin proposes the \emph{SafeMath} libraries~\cite{safemath} to help developers of Ethereum smart contracts perform proper validation on numeric inputs and prevent integer overflow and underflow vulnerabilities. 
Only two respondents in our survey mentioned that they use dependency management as a security strategy. 
Previous studies found that improper use of libraries, including security-related APIs, can introduce security vulnerabilities~\cite{ICSE-14-egele2013empirical,ICSE-15-fahl2012eve,ICSE-16-georgiev2012most,69nadi2016jumping}. Future studies could put more effort into providing \textbf{documentations of smart contract libraries with helpful examples} and \textbf{tools to facilitate library updates} for smart contract development.

To facilitate code reuse, Ethereum Virtual Machine provides an opcode, {\tt DELEGATECALL}~\cite{krupp2018teether}, for dynamically loading the bytecode of a callee contract into the caller contract at runtime. A DoS attack against the Parity wallet leveraged the vulnerability due to improper use of  {\tt DELEGATECALL}. Thus, code reuse in smart contracts can impose a higher risk than its counterpart in traditional software, highlighting the importance of \textbf{security auditing on broadly used smart contracts and libraries.}

Active maintenance ranks on top of the factors that affect the adoption of security tools for smart contracts (RQ2). Among the four tools we investigated, the most frequently used tool, \emph{Mythril}~\cite{consensys2019Mythril}, has released 102 versions since its first release on October 4, 2017 -- an average of 3 releases per month. The active maintenance would enable security tools to uncover the latest emerging security issues. As suggested in our survey, the practitioners tend to rely on security plugins in IDEs and prefer tools that cover a broad range of security issues. Thus, the future work could focus on \textbf{automatically incorporating emerging security issues into security tools} and \textbf{integrating various security tools into the IDEs.} In addition, the strategy of chasing behind the attackers is not adequate to address smart contract security. Practitioners could \textbf{proactively defense smart contracts against security attacks via external auditing and fuzzing.}

\subsection{Studies across Blockchain Platforms}
The results of RQ3 indicates that blockchain platforms impact practitioners' perceptions and practices on smart contract security. Nonetheless, previous studies usually focus on one blockchain platform. It could be interesting to investigate \textbf{the difference in security issues across different blockchains}, and \textbf{whether and how existing tools can be used across different blockchains}. In addition, practitioners of public blockchains tend to be more motivated to address smart contract security than those of consortium and private blockchains. The potential reason could be the accessibility of public blockchains to any users, even malicious attackers. Future research could investigate \textbf{whether the practitioners of consortium and private blockchains make an economic decision of security strategies based on risk assessment}.

 \section{Threats to Validity}\label{sec:threat}

\textbf{Internal Validity.} In our study, the interviewees were selected by a contact at each company or open-source project who identified the practitioners to be interviewed. The procedure partially alleviates the threat of selection bias since the interviewer has no contact with interviewees before the interviews. The threat of selection bias would always be present when the interviewees were not fully randomly sampled. However, given that our interviews include practitioners with various job roles and from different companies and open-source projects, the threat has limited effect.

As for the survey, it is possible that some of our respondents had a poor understanding of the statements for rating. Their responses may introduce noise to the data that we collected. To reduce the impact of this issue, we included an ``I don't know'' option in the survey and ignored responses marked as such. We also dropped responses that were submitted by people whose job roles are none of these: software development, testing, and project management. Two of the authors translated our survey to Chinese to ensure that respondents from China could understand our survey well. To reduce the bias of presenting the survey bilingually, we carefully translated our survey to ensure there is no ambiguity between English and Chinese terms. We polished the translation by improving clarity and understandability according to the feedback from our pilot survey.

\noindent\textbf{Construct Validity.} In our interviews, the evaluation apprehension was ameliorated by the anonymity of the interviewees, as well as the guaranty that all the information obtained during the interviews would be used only by the researchers. The interviewer might have influenced the interviewees by asking specific questions. To mitigate this risk, we used open-ended questions to elicit as much information as possible from practitioners.  The interviewees may have had a different understanding of the questions than what we had intended. To minimize this aspect, we encouraged the interviewees to ask questions at all times.

In our survey, the results are based on respondents’ self-reported responses, which may be subject to bias and not exactly represent reality. We followed recommendations to reduce social-desirability bias by ensuring respondents’ anonymity~\cite{nederhof1985methods}. The questionnaire in our survey is based on interview results instead of validated scales. Although we use factor analysis to analyze the results, it may be insufficient to validate the scales.

\noindent\textbf{Conclusion Validity.} The interviews were conducted at different locations and each interview was done in one work session. Thus, answers were not influenced by internal discussions. To ensure that the interview instrument is of high quality to obtain reliable measures, we conducted several pilots to improve the questions and layout of the interview guide prior to conducting the interviews.

In addition, we did our best to randomly select survey respondents from both companies and open-source projects. Our survey respondents come from 35 countries across six continents who are work in various job roles with a wide range of experience.

\noindent\textbf{External Validity.} To improve the generalizability of our findings regarding smart contract development, we interviewed 13 interviewees from blockchain companies and open-source blockchain projects. We further surveyed 156 respondents from 35 countries across six continents who are working for various companies or contributing to open-source blockchain/smart contract projects that are hosted on GitHub, in various job roles.
 \section{Conclusion}\label{sec:conclusion}
This work proposed a mixed qualitative and quantitative approach to explore practitioners' perceptions and practices on smart contract security. 
We recognized the disconnect between the security awareness of smart contract practitioners and the occurrence of security problems in smart contracts. We also provided practical lessons about code reuse, tool implications, and proactive defense to ensure smart contract security. Besides, we observed several differences between smart contract security and regular security: (1) Smart contract practitioners tend to have a higher security awareness than regular practitioners; (2) Smart contracts are more prone to security attacks than regular software; (3) More frequent code reuse in smart contract development imposes higher security risk than regular software development. 
Future studies could put more effort into investigating the differences in various aspects of smart contracts on top of different blockchain platforms, and generalize existing tools across different blockchain platforms. \balance
\section*{Acknowledgements}
This research was partially supported by the National Key R\&D Program of China (No. 2020YFB1005400), Australian Research Council's Discovery Early Career Researcher Award (DECRA) Funding Scheme (DE200100021), ARC Discovery Grant (DP200100020), National Science Foundation of China (No. U20A20173), Hong Kong RGC Project (No. 152193/19E), and the National Research Foundation, Singapore under its Industry Alignment Fund - Prepositioning (IAF-PP) Funding Initiative. Any opinions, findings and conclusions or recommendations expressed in this material are those of the author(s) and do not reflect the views of National Research Foundation, Singapore.

\Urlmuskip=0mu plus 1mu\relax
\bibliographystyle{abbrv}
\bibliography{smart_contract_survey}

\end{document}